# International Conference on Advanced Materials, Energy and Environmental Sustainability (ICAMEES-2018)

# Comparison of the electrochemical performance of $CeO_2$ and rare earth-based mixed metallic oxide ($Ce_{0.9}Zr_{0.1}O_2$) for supercapacitor applications

Sourav Ghosh[a,b], G. Ranga Rao[b,c*], Tiju Thomas[a,c*]

a: Department of Metallurgical and Materials Engineering, Indian Institute of Technology Madras, Chennai, 600036, Tamil Nadu, India

b: Department of Chemistry, Indian Institute of Technology Madras, Chennai, 600036, Tamil Nadu, India

c: DST Solar Energy Harnessing Centre-An Energy Consortium, Indian Institute of Technology Madras, Chennai, 600036, Tamil Nadu, India

*: Corresponding author

**Abstracts**

$CeO_2$ and $Ce_{0.9}Zr_{0.1}O_2$ are prepared from sol-gel method to investigate and compare their electrochemical properties for supercapacitor applications. Structural, morphological, and elemental studies have been done for $CeO_2$ and $Ce_{0.9}Zr_{0.1}O_2$ by XRD, SEM, and EDX. Cyclic voltammetry, galvanostatic charge-discharge, and electrochemical impedance spectroscopy techniques are used to study the electrochemical performance of these materials. Doping enhances the electrochemical performance of the electrode, by improving the specific capacitance (~150%, 243 F g$^{-1}$ from 96 F g$^{-1}$) for the doped system @2 mV s$^{-1}$ *Vs.* Ag/AgCl reference electrode in 2 mol L$^{-1}$ KOH electrolyte solution. $Ce_{0.9}Zr_{0.1}O_2$ shows only ~30% of capacitance degradation for a ten folds increase in current densities. $Ce_{0.9}Zr_{0.1}O_2$ also shows 16% capacitance degradation after 800 cycles with excellent Columbic efficiency (~100%) @2 A g$^{-1}$ current density. Partial replacement of $Ce^{4+}$ ion (0.97 Å) with $Zr^{4+}$ ion (0.84 Å) results in a decrease in lattice parameter, as confirmed by Rietveld refinement. $Ce_{0.9}Zr_{0.1}O_2$ has provided good energy, and power density of 1.128 Wh kg$^{-1}$ and 112.5 W kg$^{-1}$ respectively. Furthermore better diffusivity of the $Ce_{0.9}Zr_{0.1}O_2$ in KOH electrolyte (indicated using Randles-Sevcik equation based analysis) has been found to be correlated with better electrochemical performance. These insights presented here clearly indicate that Zr doping into $CeO_2$ results in a promising candidate material for electrochemical and supercapacitive applications.

1. Introduction

Climate change and the global need to move away from a carbon-based economy has spurred increasing interest in renewable and sustainable sources of energy. Hence increasing activity in technologies related to harnessing of solar, wind, wave energy, etc. is seen. However, renewable energy sources usually suffer from the problem of fluctuations, and hence require energy storage systems as well. This makes both battery and supercapacitor based technologies crucial for the movement towards a sustainable future.

It is now well know that high specific power and energy requirements, essential for the energy future of the globe, will require development of energy storage systems that are relevant for hybrid electrical vehicles, and portable electronics, among others [1]. Given the power requirements, it is rather evident that the development of novel materials and an improved understanding of electrochemical properties at the electrode/electrolyte interface in nano-regime would be valuable.

Based on their energy storage mechanism, supercapacitors are classified into mostly two types, electric double layer capacitor (EDLC) and pseudocapacitors [2]. Electric double layer capacitor or EDLC use ion desorption, and pseudocapacitors use surface redox reaction for storing energy. Supercapacitors are electrochemical charge storage devices known for their high power density (1-10 kW kg$^{-1}$), long cycle life (~10$^5$ cycles) and low energy density (up to 10 Wh kg$^{-1}$) [3]. Due to high power density and significantly decent energy density, Supercapacitors can be used in a hybrid system alongside battery or individually by replacing a battery, depending on the requirement of the system. Recent advancement in nanotechnology and understanding of the charge-storage mechanism in nanoscale provided new insight into supercapacitor technology. The quest for supercapacitor electrode materials other than carbon has resulted in the emergence of transition metal-based oxides as a reasonable alternative [4-7]. Rare earth metal-based oxides, especially ceria is mostly used as a catalyst, solid oxide fuel cells, luminescence, ceramics, etc. for an affordable cost, environment-friendly nature, oxygen vacancy, and variable valence state of cerium. The presence of mixed valance state and oxygen vacancy can potentially become useful for supercapacitor applications as well. Mostly because of that, recently a few works have looked into ceria and ceria-based composites as a prospective electrode material for supercapacitors, such as cerium oxide nanoparticles [8], NiO-$CeO_2$ binary alloy [9], Ce-MOF/GO [10], $CeO_2$ nanoparticles/graphene nanocomposite [11, 12], etc. These previously mentioned works have mostly used carbon material, especially graphene-based composite, to improve the electrochemical performance of Ce based materials.

Here the authors explore the possibility of rare earth metal (Ce) based mixed metallic oxide system ($Ce_{0.9}Zr_{0.1}O_2$) for supercapacitor electrode applications. Zr doping improves the thermal stability, mobility of lattice oxygen, and oxygen vacancy in ceria [13, 14]. Oxygen vacancy causes in-situ redox reaction ($Ce^{4+}/Ce^{3+}$). Increasing redox reactions on the electrode/electrolyte surface can improve the capacitance value. In the composite, lattice contraction occurs due to the partial replacement of relatively larger cerium ion by smaller Zr ion, which facilitates the oxygen vacancy in this case. Firstly, the XRD are compared to study the crystallite size and lattice constants to quantify the structural effect of doping. A morphological study by scanning electron microscopy (SEM) provides information regarding the morphological advantages of Zr doped system. Rietveld refinement and EDS are used to understand the doping concentration of the mixed metallic system ($Ce_{0.9}Zr_{0.1}O_2$). Finally, from the electrochemical studies, the superior performance of $Ce_{0.9}Zr_{0.1}O_2$ when compared to $CeO_2$ is demonstrated. A reasonable energy density (1.128 Wh $kg^{-1}$) and power density (112.5 W $kg^{-1}$) are observed for the rear earth based mixed metallic system.

## 2. Experimental section

### 2.1. Materials

Cerous nitrate hexahydrate extrapure (SRL, India), Zirconium oxynitrate hydrate exiplus (SRL, India), Citric acid (anhydrous)(Fisher scientific), Potassium hydroxide (85%) pellets (Avra synthesis, India), Sodium hydroxide

(98%) pellets (Avra synthesis, India), Ethanol are purchased and used as it is. De ionized water (18.2 mΩ) is used to prepare sodium hydroxide and potassium hydroxide solution.

### 2.2. Material synthesis

$CeO_2$ (CNP) and $Ce_{0.9}Zr_{0.1}O_2$ (CeZ) are synthesized by the sol-gel method. To prepare $Ce_{0.9}Zr_{0.1}O_2$ (CeZ), the calculated amount of $Zr(NO_3)_4, xH_2O$ and $Ce(NO_3), 6H_2O$ were added in ethanol to prepare 0.2 mol $L^{-1}$ solution to maintain the theoretical molar ratio of 10% for Zr in the composite. 0.4 mol $L^{-1}$ citric acid and ethanol solution are prepared in parallel. Both of the solutions are mixed in an equal volumetric ratio. The pH of the solution is maintained to 7 by adding 0.3 mol $L^{-1}$ NaOH solution prepared in deionized water. Then the mixed solution is heated at 80 ºC for several hrs up to gel formation. The gel is dried for overnight at 80 ºC before it is calcined at 650 ºC for 4hrs in a muffle furnace. To prepare $CeO_2$ (CNP), the same procedure is followed except the addition of Zirconium oxynitrate hydrate. After the furnace is cooled down to room temperature, the calcined substance is collected and ground to obtain a fine powder of respective materials [13, 15].

### 2.3. Material characterization

For obtaining the crystallographic information, powder X-ray diffraction measurements are performed with a Bruker AXS D8 Advance diffractometer by employing Cu Kα (1.5406 Å). The analysis is operated at 40 kV and 30 mA at room temperature for 5º to 80º with a step size of 0.03º. Scanning electron microscopy (SEM) images and Energy dispersive spectroscopy (EDS) are taken by using FEI Quanta FEG 400 and Bruker X Flash 6I 10 to study morphological and elemental analysis.

### 2.4. Electrochemical characterization

The electrochemical measurements are done in a CHI 7081C electrochemical workstation. A three-electrode based system is obtained with Ag/AgCl as the reference electrode, Platinum foil (1 × 2 $cm^2$) as the counter electrode and 2 mol $L^{-1}$ KOH solution. 85 wt% of synthesized metal oxide powder and 10 wt% acetylene black are mixed in an agate mortar. 5 wt% Polyvinylidene fluoride (PVDF) dissolved in N-Methyl-2-pyrrolidone

(NMP) is added in the mixed powder to make the slurry of active material. The active material is then coated on a Nickel (Ni) foil (1 cm × 5 cm) on a 1 cm² area. The coated electrode is dried overnight in a vacuum oven for 60 ºC. Cyclic voltammetry (CV), galvanostatic charge-discharge (GCD) and electron impendence spectroscopy (EIS) measurements are carried out to study electrochemical properties of the as-synthesized materials.

### 2.5. Results and discussion

**2.5.1 Structural, morphological and elemental analysis**

The powder X-ray diffraction spectrum of CNP and CeZ are given in Fig. 1 with corresponding peaks. The peaks at 28.5º, 33.1º, 47.5º, 56.3º, 59.1º, 69.4º, 76.7º, 79.1º, 88.5º in Fig. 1(a) are assigned to (111), (200), (220), (311), (222), (400), (331), (420) and (422) respectively. These peaks are in agreement with cubic fluorite (CaF2 structure) crystal structure of $CeO_2$ as given in ICDD: 98-007-2155. The peaks in Fig. 1(b) are assigned to the same set of planes, but the angle of peaks are slightly shifted to 28.8º, 33.3º, 47.8º, 56.7º, 59.4º, 69.8º, 77.1º, 79.5º, and 88.9º and indexed as same cubic fluorite structure ( ICDD: 98-010-4708). Interestingly the peaks intended to Zirconium oxide ($ZrO_2$) are absent in the powder XRD spectrum, which indicates at successful incorporation of Zr into $CeO_2$ lattice structure [14, 16].

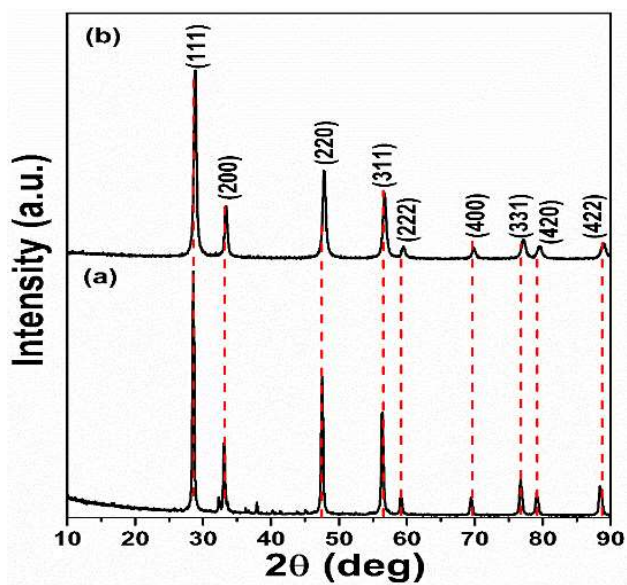

Figure 1 Powder X-ray diffraction spectrum for (a) $CeO_2$, and (b) $Ce_{0.9}Zr_{0.1}O_2$

The average crystallite size is calculated from the Scherrer equation as given below:

$$d = \frac{K\lambda}{\beta \cos\theta}$$

In the above equation d denotes the crystallite size, K is shape factor of value 0.9, $\lambda$ is the wavelength of X-ray, $\beta$ is full width half maximum (FWHM), and $\theta$ is the value of Bragg angle. The average crystallite size calculated by the above equation, corresponding to all the peaks given in Fig.1 is 23.76 nm and 15.05 nm for CNP and CeZ respectively.

The lattice parameter of CNP (5.4109 Å) and CeZ (5.3992 Å) can be seen in Table 1 and agree with the literature [14,15]. The lattice parameter of the CNP and CeZ is confirmed by Rietveld refinement, as shown in Fig. 2. The values of parameters corresponding to Rietveld refinement are given in Table 1.

**Table 1.** Structural parameters as deduced from Rietveld refinement for $CeO_2$ and $Ce_{0.9}Zr_{0.1}O_2$

| Parameters | CNP | CeZ |
| --- | --- | --- |
| Space group | Fm-3m | Fm-3m |
| Lattice parameter | 5.4108 Å | 5.3992 Å |
| Average bond length | 2.3430 Å | 2.3379 Å |
| $R_p$ | 10.1 | 5.28 |
| $R_{wp}$ | 14.1 | 7.15 |
| $\chi^2$ | 3.92 | 1.09 |

The structural parameter values remain in agreement with reported results in the literature [13, 16, 17]. As seen in Fig. 1 the peaks are shifted towards higher 2Θ values in case of CeZ as well as the lattice parameter value has decreased. This phenomenon of peak shifting and lattice contraction can be assigned to the inclusion of $Zr^{4+}$ in cubic fluorite $CeO_2$ crystal structure and partial replacement of $Ce^{4+}$ ion (0.97 Å) with $Zr^{4+}$ ion (0.84 Å) [14, 16,17]. The incorporation process of $Zr^{4+}$ ion into $CeO_2$ can be seen clearly in Fig.2. This doping process is likely to result in a significant positive charge generation and an increase in the electrical conductivity of the material. Increased conductivity will provide a better current-voltage response, which will, in turn, result in better supercapacitive performance for CeZ [18]. Apart from this, Rietveld refinement result provided information on the chemical content of the unit cell as well (8.0000 O + 0.4184 Zr + 3.7453 Ce). The ratio of Ce to Zr content according to Rietveld refined information is 9:1, which denotes the stoichiometric composition of the bimetallic oxide is maintained properly as targeted.

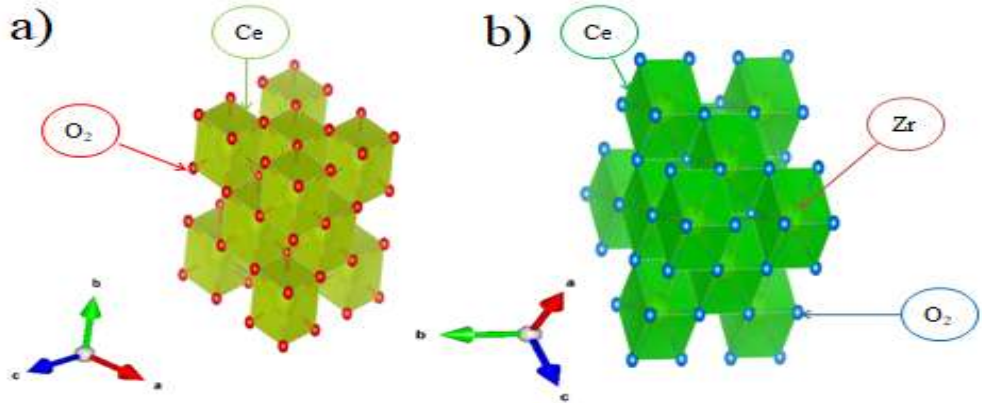

Figure 2 Pictorial representation of (a) $CeO_2$, and (b) $Ce_{0.9}Zr_{0.1}O_2$ packing diagram generated by Rietveld parameters by using VESTA.

Morphological study of the as-prepared materials has been done from the SEM images of the samples as given in Fig.3. SEM image of CeZ clearly shows spherical nanoparticles where CNP has no such significant structure. Mainly shaped and well-defined morphology can improve ion transport as an advantage for supercapacitor application. Distinct morphological structure and shape also helps to relax the material and hold structural integrity against volume expansion during continuous charging and discharging process. Thus, it takes a significant role in cycling stability and long term performance of the supercapacitor as well. Fig. 4 shows the EDS spectrum of CeZ, which proves the presence of Zr in the CeZ system. Besides confirming the successful doping of Zr in Ceria, EDS also provides approximate information on the stoichiometrical composition of the binary metallic system. From the atomic percentage, as provided by the EDS spectrum, the doping concentration can be confirmed as approximately 10% ($Ce_{0.93}Zr_{0.07}O_2$).

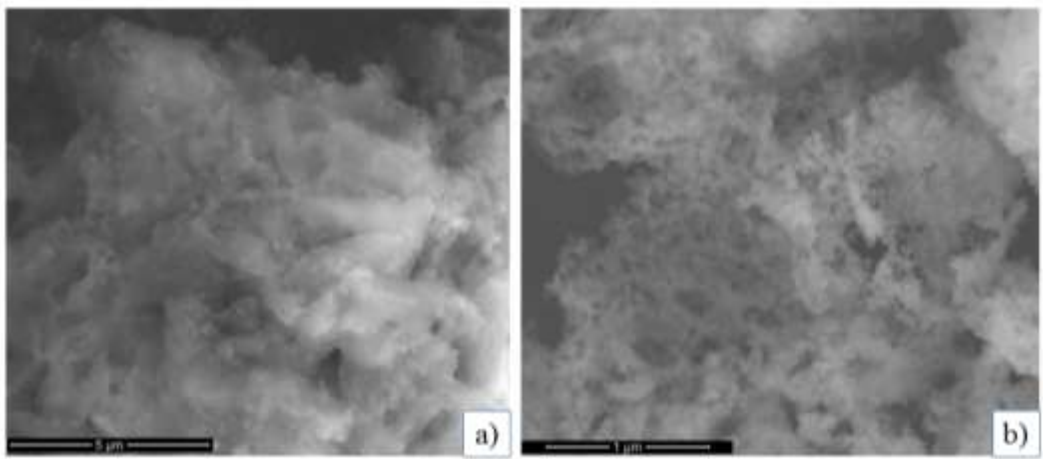

Figure 3 SEM image of (a) $CeO_2$, (b) $Ce_{0.9}Zr_{0.1}O_2$

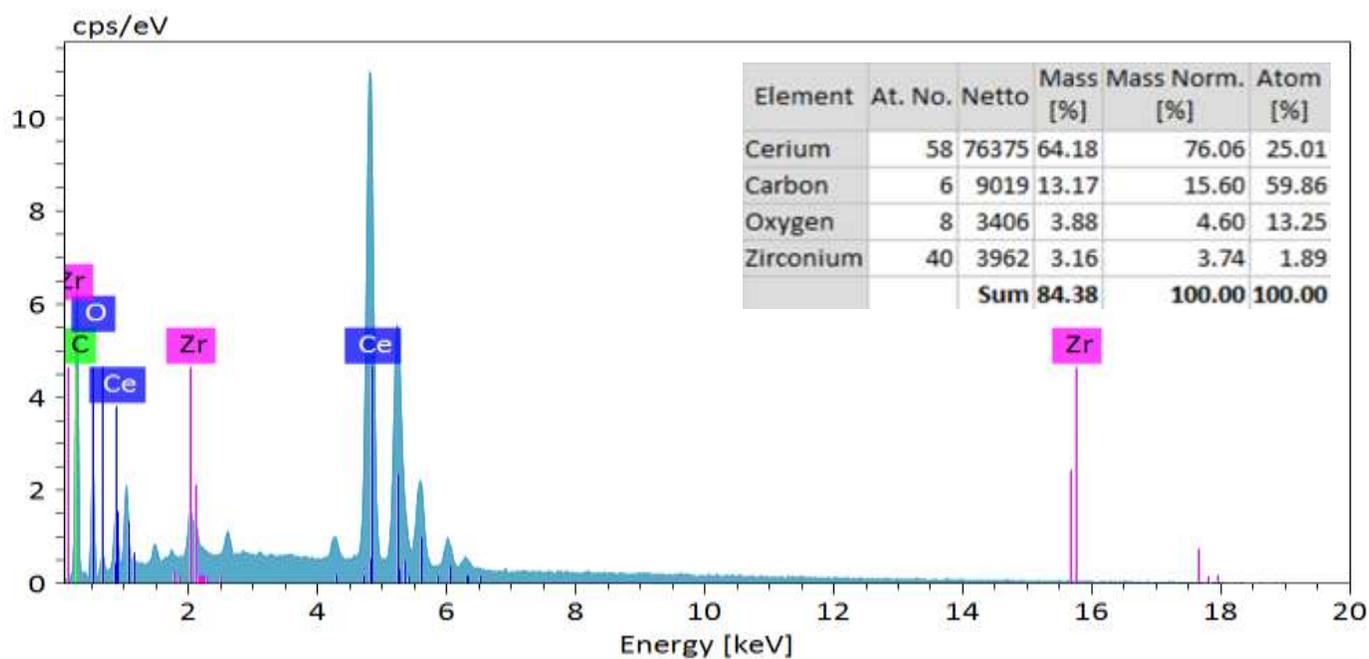

Figure 4 EDS spectrum of $Ce_{0.9}Zr_{0.1}O_2$. Atomic percentage table is provided for the spectrum in the inset.

### 2.5.2. Electrochemical characterization

Cyclic voltammetry (CV) diagram for three-electrode system is depicted in Fig. 5a and 5b. The area under the curve is associated with the value of specific capacitances. The maximum value of specific capacitance calculated for CNP is 96 F g$^{-1}$ @ 2mV s$^{-1}$ whereas for CeZ it is 233 F g$^{-1}$ for the same condition. The range of scan rates is less for CNP as the resultant voltammetric capacitance becomes much lower after that. The incorporation of Zr has caused an increase in both electronic and ionic conductivity of CeZ compared to CNP. Thus, improved the reaction kinetics and enhanced electrode utilization, which is favorable for supercapacitor application as observed from the huge increase (~150%) in specific capacitance for CeZ at 2 mV s$^{-1}$. CV curve of both CNP and CeZ show redox peaks, which indicates at their pseudocapacitive behavior. Interestingly, the oxidation and reduction peaks are almost constant concerning the X-axis, i.e., the potential for CNP. That indicates the redox reaction is occurring at the same redox potential, which denotes the presence of surface adsorbed species and reversible electrochemical nature of ceria. However, for CeZ the distance between oxidation and redox peaks are slowly increasing with increasing scan rates. This phenomenon is indicative of quasi-reversible nature of the electrochemical reaction [18,19]. The quasi-reversible reaction is probably due to the incomplete cyclic transition of $Ce^{3+}$ and $Ce^{4+}$ ions in high scan rates. Also, the specific capacitance value decreases with increasing scan rate. At higher scan rates, the internal structure of material becomes less accessible to the ions, which results in low specific capacitance, as shown in Fig. 5c.

The galvanostatic charge-discharge profile of CNP and CeZ can be observed in Fig. 5d and 5e, respectively. The charge-discharge profile has been noted for a significant variation of current density (0.5 A g$^{-1}$ to 10 A g$^{-1}$) for both CNP and CeZ. CeZ shows a maximum specific capacitance of 40 F g$^{-1}$ at 0.5 A g$^{-1}$ in comparison to 22.3 F g$^{-1}$ of CNP in same current density. Thus Zr doping has ensured 80% improvement of specific capacitance calculated from the charge-discharge diagram. The specific capacitance decreases with increasing current densities (Fig. 5f) is because the inner active sites of the material become less approachable for ions for electrolytes, in this case, KOH solution. Non-triangular shaped galvanostatic discharge curves show the presence of plateaus in the ~0.3 V region for both CNP and CeZ as shown in Fig. 5d, and 5e. The presence of plateau in the discharged area of the galvanostatic charge-discharge curve is typical of pseudocapacitors. This result is in agreement with CV behavior as well. It can be noted that the charged and discharged voltage is not equal for CeZ in higher current densities. This charge-discharge behavior can be due to the quasi-reversible electrochemical reaction as confirmed from the current-voltage response previously.

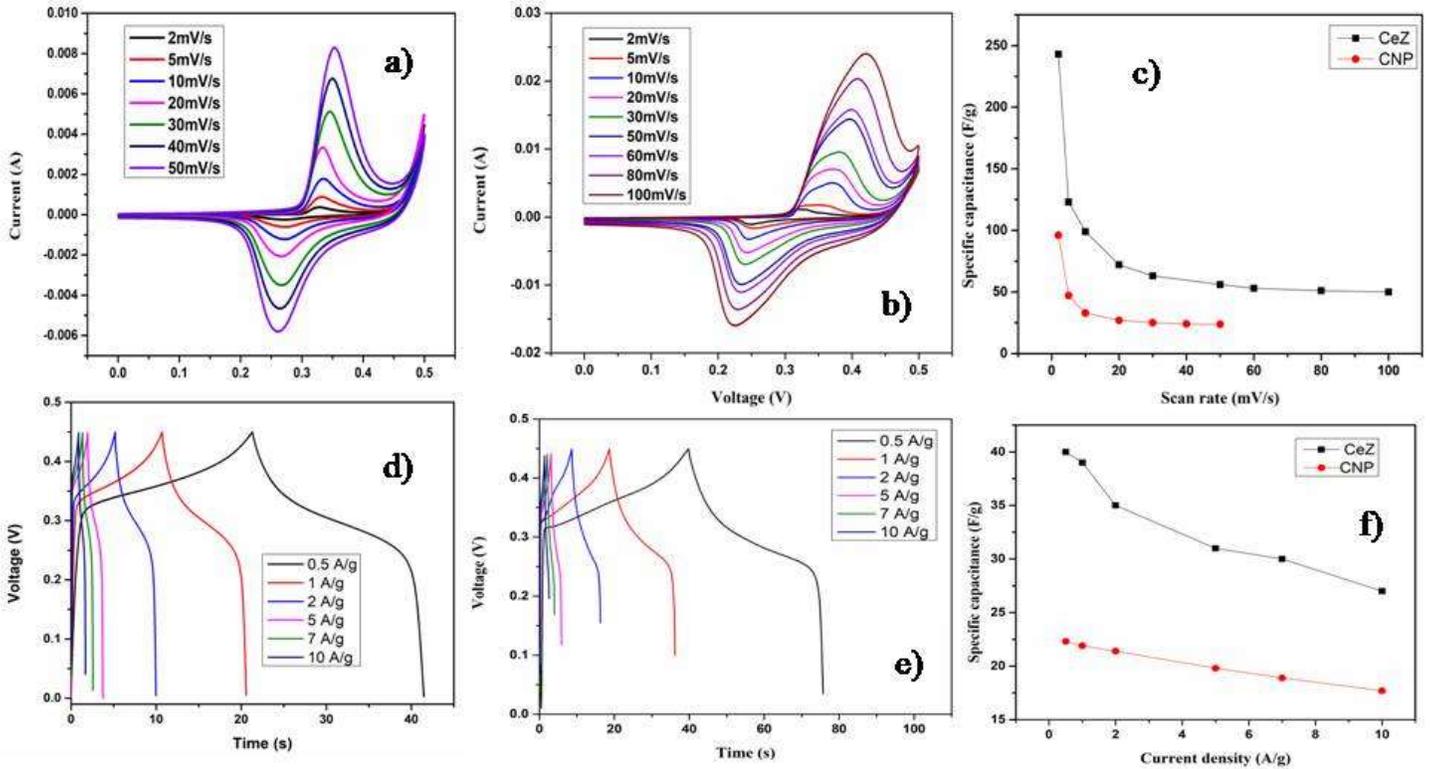

Figure 5 Cyclic voltammetry diagrams of (a) $CeO_2$, (b) $Ce_{0.9}Zr_{0.1}O_2$ (c) comparison of specific capacitance with increasing scan rates between $Ce_{0.9}Zr_{0.1}O_2$ and $CeO_2$ *Vs.* Ag/AgCl reference electrode in 2 mol $L^{-1}$ KOH electrolyte. Galvanostatic charge-discharge characteristics of (d) $CeO_2$, and (e) $Ce_{0.9}Zr_{0.1}O_2$, (f) comparison of specific capacitance with an increasing current density between $Ce_{0.9}Zr_{0.1}O_2$ and $CeO_2$ *Vs.* Ag/AgCl reference electrode in 2 mol $L^{-1}$ KOH electrolyte.

The overall symmetric nature of the curves (Fig. 5d-e) confirms the high columbic efficiency of the supercapacitors (~100%). High columbic efficiency ensures the high energy efficiency of the energy storage system, which is a necessity of modern age technologies. Besides energy efficiency, another important set of parameters for an electrical energy storage system are energy density and power density. Energy density and power density are calculated based on the discharge curves presented in Fig. 5d-e. At a current density of 0.5 $Ag^{-1}$, CeZ shows a decent energy density and power density of 1.128 Wh $kg^{-1}$ and 112.5 W $kg^{-1}$ compared to 0.678 Wh $kg^{-1}$ and 112.45 W $kg^{-1}$ of CNP respectively. This proves Zr doped ceria as a promising material for supercapacitor application and demands further investigation and modification.

One of the prime factors for long cycling life and specific energy delivery for an electrical energy storage device is the resistance of the cell. This total cell resistance comprises of the i) electrode resistance, ii) electrode and electrolyte interface resistance, iii) resistance due to ion diffusion in the electrolyte especially through the porous structure, and iv) charge-transfer resistance. As a matter of fact, due to an alternating current (AC), the capacitive component remains ever-present in an electrochemical system. Because of that, AC impedance analysis provides useful information regarding the power capability of an electrical device such as a supercapacitor. Frequency response in the range of 0.01 Hz to 100 kHz of CNP and CeZ has been studied by Electron impedance spectroscopy (EIS) technique. Fig. 6a to 6c is plotted based on the EIS data. From the Nyquist plot of Fig.6a, it can be deduced that both of the material has similar frequency response for low-frequency zone, but the impedance nature of CeZ increases in high-frequency zone. This high impedance nature can be confirmed from equivalent circuit modeling as given in Fig. 6b. Rs represents series resistance, C denotes the capacitance value of the circuit, Rct symbolizes charge transfer resistance, and W shows the Warburg impedance which appears due to the diffusion-controlled nature of the electrochemical cell, which is typical of metal oxide based supercapacitors. The equivalent circuit has been drawn and fitted by Zsimpwin V.3.21 software. The series resistant value (Rs) increases from 1.398 Ω to 1.581 Ω for Zr doped material. However, both have similar charge transfer resistance (Rct) of 0.001 Ω, which indicates; the conductive mechanism is identical for both of the materials. The capacitance value C is mostly double layer capacitance, due to the ion adsorption on the electrode-electrolyte surface [19, 20, 21]. The higher ion accumulation on the electrode/electrolyte surface is because of the significant contribution of double layer capacitance in total specific capacitance. This double layer contribution is more for CNP compared to CeZ. Increase in active surface redox reaction as an effect of doping in case of CeZ might result in higher specific capacitance as well the lower contribution of double layer capacitance.

Table 2 Equivalent circuit parameters as deduced from Nyquist plot for $CeO_2$ and $Ce_{0.9}Zr_{0.1}O_2$.

| Parameters | CNP | CeZ |
|---|---|---|
| $R_s$ (Ω) | 1.398 | 1.581 |
| $R_{ct}$ (Ω) | 0.001 | 0.001 |
| C (F) | 0.0009 | 0.0005 |
| W (Ω) | 0.0002 | 0.0003 |
| Error | 0.1020 | 0.1101 |

From Fig. 7a the time constant (t0) can be calculated from the value of f0, which is maximum peak value of C" calculated from the below equation [18, 21]:

C"(ω) is the imaginary part of the capacitance value, Z'(ω) is the real part of impedance, and Z(ω) is the absolute impedance. The calculated t0 is higher for CNP ($9.51 \times 10^{-3}$) compared to CeZ ($8.34 \times 10^{-3}$), as CeZ has a higher value of $f_0$ (Fig. 7a). The low value of the time constant is because of the well-defined spherical structure of CeZ, which has improved the ion accessibility of the material in time of the charge-discharge process. Higher ion accessibility of ions into nanostructure of the material enhances the ion diffusion process in electrode/electrolyte surface.

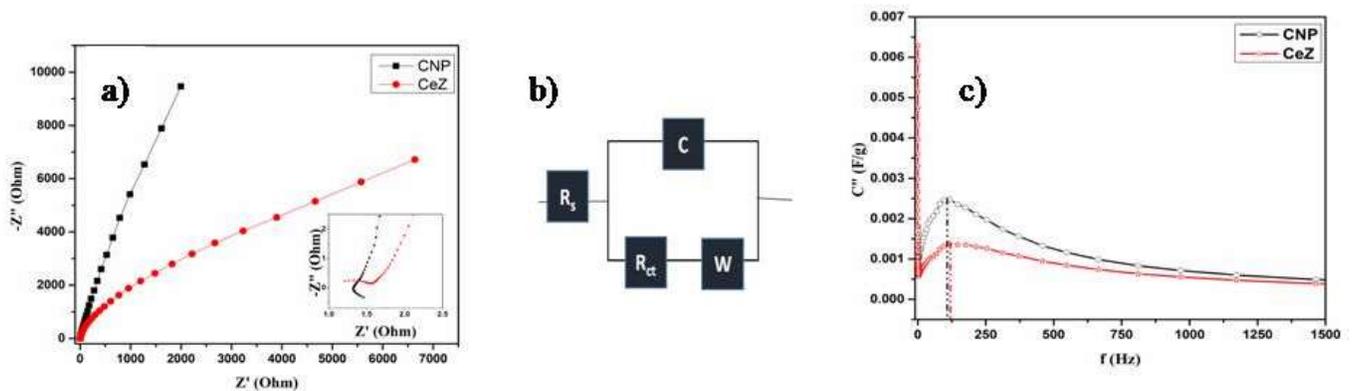

Figure 6 (a) Nyquist plot, (b) equivalent circuit of supercapacitor system, (c) Plot of frequency vs. imaginary part of the capacitance.

For the better understanding of the diffusion mechanism in the electrochemical process, the peak current vs. square root of scan rate has been plotted as per the Randles-Sevcik equation (Fig. 12). The relevant equation is,

$$\frac{I'}{\upsilon^{\frac{1}{2}}} = (2.69 \times 10^5) n^{\frac{3}{2}} S D^{\frac{1}{2}} C'$$

In the above equation, I' is the peak current (A), $\upsilon$ is the scan rate (V s$^{-1/2}$), n is the number of electrons lost or gained in the process, S is the surface area of the electrode (cm$^2$), D is the diffusivity of the reactant (cm$^2$ s$^{-1}$) and C' is the initial reactant concentration (mol cm$^{-3}$). So, it is evident that the diffusivity can be understood from the slope of the plot and in this case, CeZ is more diffusive in electrolyte compared to CNP, which is one of the reasons for better supercapacitive performance. Although the peak current vs. $\upsilon^{1/2}$ is almost linear as it should be in case of the freely diffusive system the modest shift from linearity especially in case of CeZ demonstrates, i) surface redox reaction by surface adsorbed species, and ii) quasi-reversibility of the electrochemical reaction [18, 19]. This argument is supported by CV and EIS results as well.

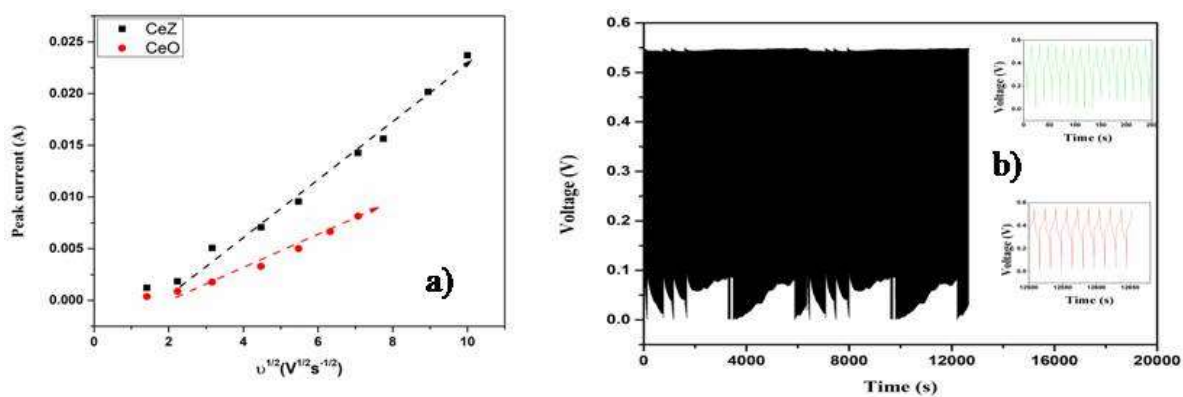

Figure 7 (a) Relation between the peak current to the square root of scan rate of Ce$_{0.9}$Zr$_{0.1}$O$_2$ and CeO$_2$, (b) 800 cycle run of Ce$_{0.9}$Zr$_{0.1}$O$_2$ with starting and ending segments present inset.

As a part of material testing for supercapacitor, CeZ has been tested for 800 cycles, (Fig.7b) at a current density of 2 A g$^{-1}$. The cyclic stability test shows 16% capacitance degradation in comparison with the first cycle. However, the binary metal oxide maintains a tremendous columbic efficiency of 100%, which reaffirms it as an energy-efficient system. The flower-like structure might have relaxed the aftermath of structural damage during the cycling process as well as relaxed the effect of volume expansion to hold on to the structural integrity of the material nanostructure. The lower time constant, good cyclic stability, high diffusivity, fantastic coulombic efficiency along with high specific capacitance promotes CeZ as a promising candidate for supercapacitor electrode material.

## 3. Conclusion

A transition metal (Zr) has been incorporated into a rare earth metal oxide ($CeO_2$) to synthesize mixed metallic oxide ($Ce_{0.9}Zr_{0.1}O_2$). The binary metal oxide is investigated to study their electrochemical properties and suitability for supercapacitor application. The mixed metallic oxide has provided 243 F g$^{-1}$ of specific capacitance at a scan rate of 2 mV s$^{-1}$ with specific energy, and power density of 1.128 Wh kg$^{-1}$ and 112.5 W kg$^{-1}$ respectively. An improved time constant, cyclic stability, diffusivity, and the higher specific capacitance value of CeZ over pure $CeO_2$ prove rare earth-based mixed metallic oxide ($Ce_{0.9}Zr_{0.1}O_2$) as an encouraging material for supercapacitors. However, the specific capacitance value is reasonably low compared to other transition metal oxides as reported. Hence, in spite of the promising result, further research is needed to improve the electrochemical performances to promote this material to devise standard.

## 4. Acknowledgment

Sourav Ghosh acknowledges the support from the HTRA fellowship for research scholars of IIT Madras with heartfelt gratitude. Authors thank the Department of Metallurgical and Materials Engineering, Department of Chemistry, and Indian Solar Energy Harnessing Centre, Indian Institute of Technology Madras. Authors gratefully acknowledge funding from Department of Science and Technology, India (MET1617146DSTXTIJU, DST/TMD/SERI/HUB/1(C)/SOL1819001DSTXHOCX) and Ministry of Electronics and Information Technology, India (ELE1819353MEITNAK) for this work.